\pdfoutput=1
\documentclass[
superscriptaddress,
preprintnumbers,
amsmath,amssymb,
longbibliography
]{revtex4-1}

\usepackage{graphicx}% Include figure files
\usepackage{dcolumn}% Align table columns on decimal point
\usepackage{bm}% bold math
\usepackage{color}
\usepackage{hyperref}
\usepackage{xcolor}
\usepackage{braket} % for braket notation
\usepackage{cancel}
\usepackage{comment}
\usepackage[nodisplayskipstretch]{setspace}
\usepackage[skip=8pt]{parskip}
\usepackage{setspace}
\allowdisplaybreaks

\usepackage[raggedright]{titlesec}
\setcitestyle{round}

\expandafter\def\expandafter\normalsize\expandafter{%
  \normalsize  
  \setlength\abovedisplayskip{4ex}
  \setlength\belowdisplayskip{4ex}
  \setlength\abovedisplayshortskip{4ex}
  \setlength\belowdisplayshortskip{4ex}
}

\newcommand{\mg}[1]{\lvert {#1} \rvert}

\begin{document}

%\preprint{APS/123-QED}

\title{Strongly correlated multi-electron bunches from interaction with quantum light\\\textbf{Quantum light produces many-electron correlations}}% Force line breaks with \\

\author{Suraj Kumar$^{\dagger}$}
\affiliation{
School of Electrical and Electronic Engineering, Nanyang Technological University, 50 Nanyang Avenue, Singapore 639798, Singapore
}

\author{Jeremy Lim$^{\dagger}$}
\affiliation{%
Science, Mathematics and Technology, Singapore University of Technology and Design, 8 Somapah Road, Singapore 487372, Singapore
}%

\author{Nicholas Rivera}
\affiliation{Department of Physics, Harvard University, Cambridge MA, 02138}

\author{Wesley Wong}
\affiliation{
School of Electrical and Electronic Engineering, Nanyang Technological University, 50 Nanyang Avenue, Singapore 639798, Singapore
}

\author{Yee Sin Ang}
%\email{yeesin\_ang@sutd.edu.sg}
\affiliation{%
Science, Mathematics and Technology, Singapore University of Technology and Design, 8 Somapah Road, Singapore 487372, Singapore
}%

\author{Lay Kee Ang}
%\email{ricky\_ang@sutd.edu.sg}
\affiliation{%
Science, Mathematics and Technology, Singapore University of Technology and Design, 8 Somapah Road, Singapore 487372, Singapore
}%

\author{Liang Jie Wong}
\email{liangjie.wong@ntu.edu.sg}
\affiliation{
School of Electrical and Electronic Engineering, Nanyang Technological University, 50 Nanyang Avenue, Singapore 639798, Singapore
}

%\date{\today}

{
\let\clearpage\relax
\maketitle
}

\textbf{Abstract: Strongly correlated electron systems are a cornerstone of modern physics, being responsible for groundbreaking phenomena from superconducting magnets to quantum computing. In most cases, correlations in electrons arise exclusively due to Coulomb interactions. In this work, we reveal that free electrons interacting simultaneously with a light field can become highly correlated via mechanisms beyond Coulomb interactions. In the case of two electrons, the resulting Pearson correlation coefficient (PCC) for the joint probability distribution of the output electron energies is enhanced over 13 orders of magnitude compared to that of electrons interacting with the light field in succession (one after another). These highly correlated electrons are the result of momentum and energy exchange between the participating electrons via the external quantum light field. Our findings pave the way to the creation and control of highly correlated free electrons for applications including quantum information and ultra-fast imaging.}

\textbf{Teaser: Many electrons interacting with quantum light can become highly correlated in energy even in the absence of coulomb repulsion}

%================================================================
% SECTION: INTRODUCTION
%================================================================
The interaction of free electrons with light forms the basis of important imaging paradigms including cathodoluminescence (CL) microscopy~\cite{GarciaDeAbajo2010, DiGiulio2021, Liebtrau2021, GarciaDeAbajo2021}, electron energy-loss spectroscopy (EELS)~\cite{VERBEECK2004207, GarciaDeAbajo2010, Verbeeck2010, VanTendeloo2012, EGOAVIL20141, Krehl2018, Polman2019, Liu2019, Liebtrau2021, GarciaDeAbajo2021, lopez2021}, and, photon-induced near-field microscopy (PINEM)~\cite{Barkwick2009, Park2010, Feist2015, Piazza2015, Priebe2017, vanacore2018, Liu2019, Wang2020, Harvey2020, Dahan2020, Liebtrau2021, GarciaDeAbajo2021, Dahan2021, Shiloh2022, Jeremy2023}. In addition to revealing the most fleeting dynamics of matter and radiation down to subatomic length scales, free electron-light interactions have also been leveraged as a versatile method for shaping both the output free electron wavefunction~\cite{Barkwick2009, Park2010, Kaminer2015a, Feist2015, Piazza2015, Priebe2017, vanacore2018, Liu2019, Wang2020, Harvey2020, Liebtrau2021, Vanacore2020, Yalunin2021, Feist2015, Zhao2021b, Morimoto2018, LJ2015, Priebe2017, Kozak2018, Kozak2018b, Lim2019, Harris2015, Kaminer2015, vanacore2018, Grillo2014, Vanacore2019, Jones2016} and the emitted radiation~\cite{Faresab2015, Talebi2016,  Guzzinati2017, Roques2018, Gover2018, Pan2018, DiGiulio2019, Pan2019, Gover2019, Kfir2019, Kfir2020, Karnieli2021, Karnieli2021b, DiGiulio2021, Kfir2021, Wong2021, Wong22, Baranes2022, Wong_Kaminer_2021, Carbone2012}.

Recent works have shown that it is possible to induce electron-electron entanglement~\cite{Kfir2019}, realize non-trivial quantum states of light (e.g., squeezed states, displaced Fock states)~\cite{BenHayun2021}, use exchange mechanisms between free electrons to possibly create single attosecond electron pulses~\cite{Talebi_Březinová_2021}, and entangle distant photons~\cite{Baranes2022}  by subjecting photonic states to multiple successive interactions with quantum electron wavepackets (QEWs) in quantum PINEM (QPINEM) setups. It should be emphasized that in these cases, each QPINEM interaction involves only a single-electron QEW. Recently, correlations between multiple QEWs arising as a result of Coulomb interactions have been explored theoretically and experimentally\cite{haindl22, Hommelhoff22}. However, Coulomb interactions leads to only one specific type of correlation in which the electrons repel each other in physical space as much as possible. The question arises as to whether there might be complementary ways of tailoring correlations between multiple free electrons.

Here, we show that strong correlations between multiple QEWs can be created and tailored when multiple QEWs interact with a photon field -- even in the absence of Coulomb interactions. Instead, the quantized energy of the external photon field unexpectedly becomes a means of inter-particle interaction between the individual electrons. The individual electrons can thus trade quanta of momentum and energy with each other even when they are too far-spaced for Coulomb interactions to matter -- we refer to this phenomenon as simply "inter-electron momentum exchange" in the rest of this paper for brevity. We show that inter-electron momentum exchange results in output electrons that are highly correlated in energy. Specifically, we show that the Pearson Correlation Coefficient (PCC) of the participating electrons is enhanced by over 13 orders of magnitude compared to the correlation that already exists from the free electrons interacting with the photon mode in succession (one after another) where PCC is a measure of linear correlation between the two electrons which takes on values between -1 and 1. A PCC value close to 1 indicates strong linear correlation between two quantities i.e., increasing one also increases the other. A PCC value close to -1 on the other hand indicates that increasing one quantity is correlated with the decrease in the other. Finally, we show that post-selecting an electron in a multi-QEW interaction with quantum light is a robust way of tailoring electron-electron correlations in energy. The electrons post-interaction exhibit varied entanglement patterns. Our findings show that inter-electron momentum exchange could provide the means to correlate and entangle  large numbers of electrons with a photon field and shape free electron correlations for applications like ultra-fast imaging and quantum information science.
\begin{figure*}[ht!]
\centering
\includegraphics[width=0.7\textwidth]{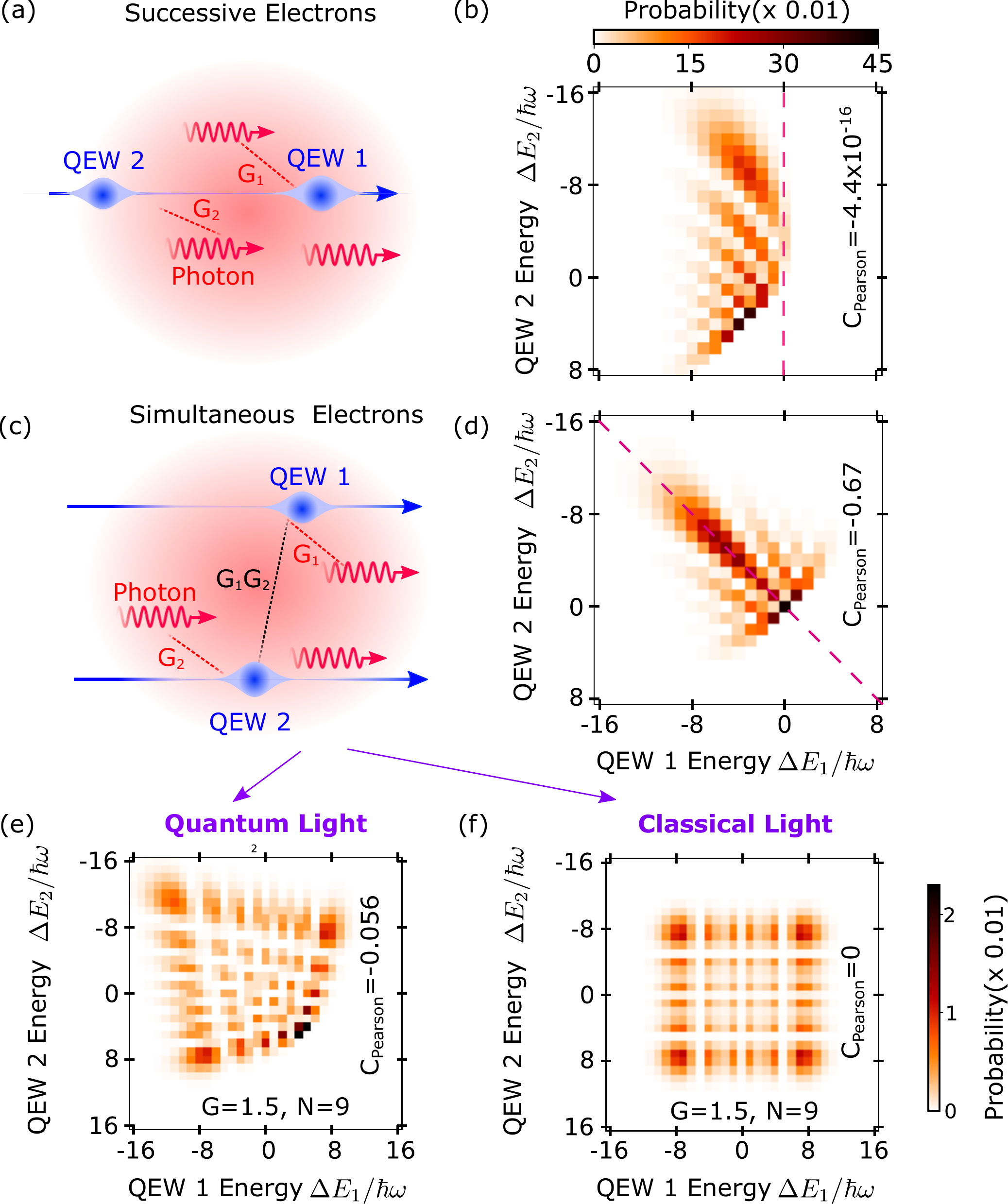}
\caption{\textbf{Simultaneous interaction of two free electron QEWs with quantum light produces highly correlated electrons.} We consider a light field prepared in the vaccum Fock state, through which two quantum electron wavepackets (QEWs) are passed. These QEWs, hereafter referred to as QEW 1 and QEW 2, couple to photons with dimensionless strengths ${G}_{1}=2$ and ${G}_{2}=2$ respectively. (a) In a quantum two-electron-light interaction without inter-electron momentum exchange, the QEWs pass through an electromagnetic environment one after the other successively. QEW 1 exchanges energy with the light field with the restriction that the final photon number be non-negative. QEW 2 then interacts with the changed light field and we measure the energy of both electrons. (b) The joint probability distribution of the energy gain of QEW 1 and QEW 2 respectively is shown here for the case without inter-electron momentum exchange. We see that because of the restriction on QEW 1, the probability is zero for any increase in its energy (beyond pink dotted line). This is because for QEW 1 to increase in energy, it must absorb a photon from the vaccum state which is not possible. The distribution is also asymmetrical and the PCC is low and is of the order $10^{-16}$. (c) In a two-electron-light quantum interaction with inter-electron momentum exchange, the two QEWs interact simultaneously with the light field. This results in the electrons exchanging energy through a non-Coulombic inter-electron "momentum exchange" term with effective dimensionless strength $|\mathcal{G}_{1}\mathcal{G}_{2}|$. The resulting absorption and emission of photons by the electrons leads to changes in the output electrons' energy joint probability distribution. (d) We compare the QEW energy gain joint distribution for the inter-electron momentum exchange case corresponding to (b). We see that the total photon number at the end of interaction, measured along the diagonals of the plot cannot be lesser than zero. We also see that the joint distribution shows symmetry along the pink dotted line. The PCC is 0.67 in magnitude and 15 orders of magnitude larger compared to the case without momentum exchange. (e),(f) We compare the correlation in energy between QEW 1 and QEW 2 for simultaneous interaction with light while (e) treating light as quantum and (f) as a classical field, showing that the phenomenon of enhanced PCC by inter-electron momentum exchange is enabled by the quantum nature of light. Here we have fixed the average photon number to be $9$ and the coupling constant to be $G=1.5$.}
\label{fig_1}
\end{figure*}
\begin{figure*}[ht!]
\centering
\includegraphics[width=1.01\textwidth]{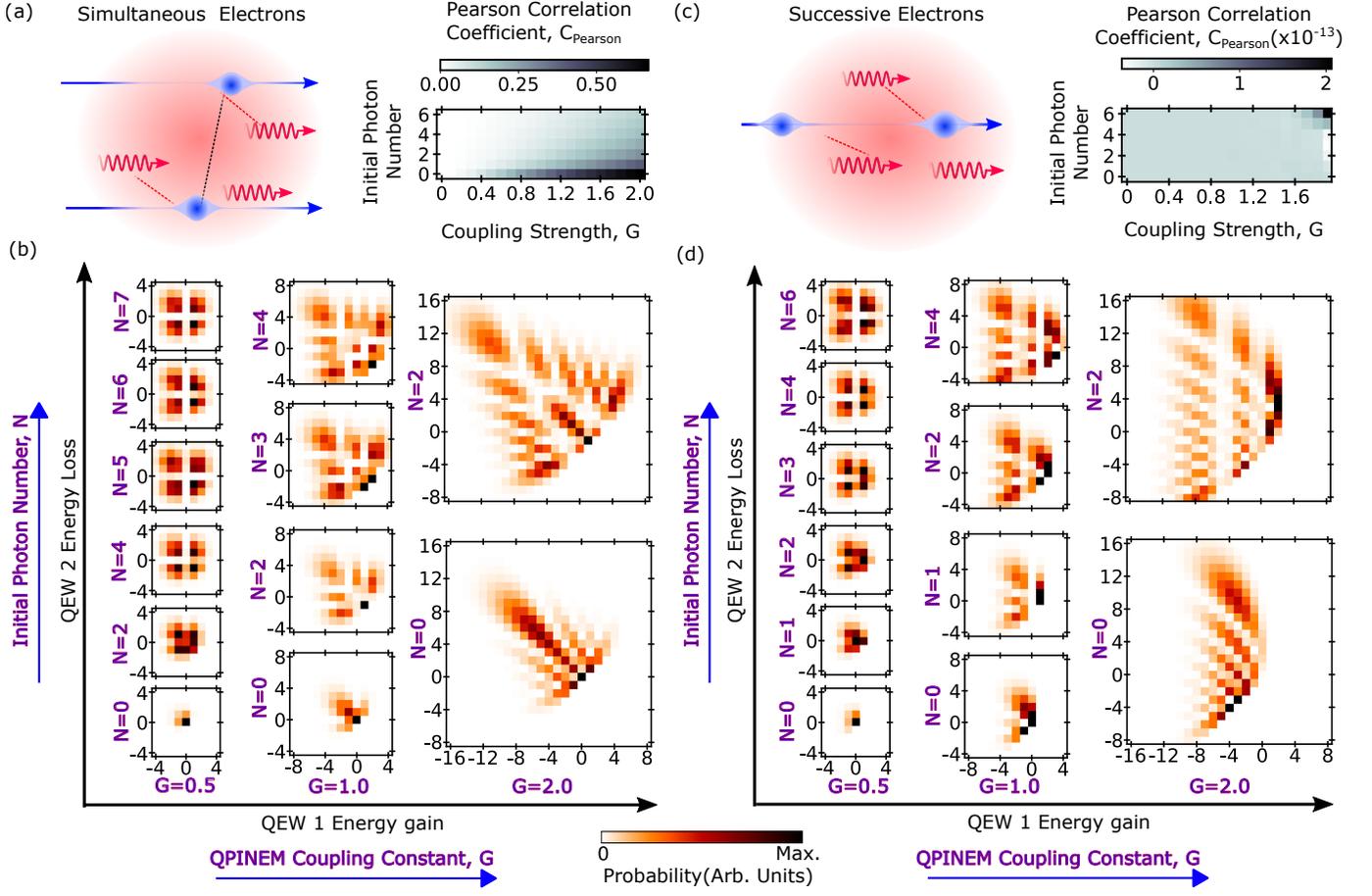}
\caption{\textbf{Stronger correlations in the output electrons are favored by lower initial photon numbers and larger interaction strengths}. (a) We show the case of simultaneous electrons with inter-electron momentum exchange and plot the colormap for the Pearson correlation coefficient with varying coupling strength $G$ and initial photon number. Note that since all PCC values here are negative we plot the absolute value. (b) We display energy correlation plots post-interaction for the case of two QEWs (Quantum Electron Wavepacket) interacting with light as we vary the interaction strength and photon number in the presence of inter-electron momentum exchange. We see that as photon number increases, especially for low interaction strength, the plots approach the classical limit and become more box-like. However for larger interaction strength it takes a much larger photon threshold for the plot to approach the classical limit. (c) For completeness, we show the case without inter-electron momentum exchange with the corresponding colormap for the PCC. (d) We observe that for low interaction strength and high photon number, the plot similar to (b) approaches the classical limit. We can thus designate a regime for which both cases tend to classical behaviour. At higher interaction strengths, the plots deviate from the classical case as expected. In general the correlation grows weaker as we increase average photon number and grows stronger as we increase interaction strength.}
\label{fig_2}
\end{figure*}
\begin{figure*}[ht!]
\centering
\includegraphics[width=0.7\textwidth]{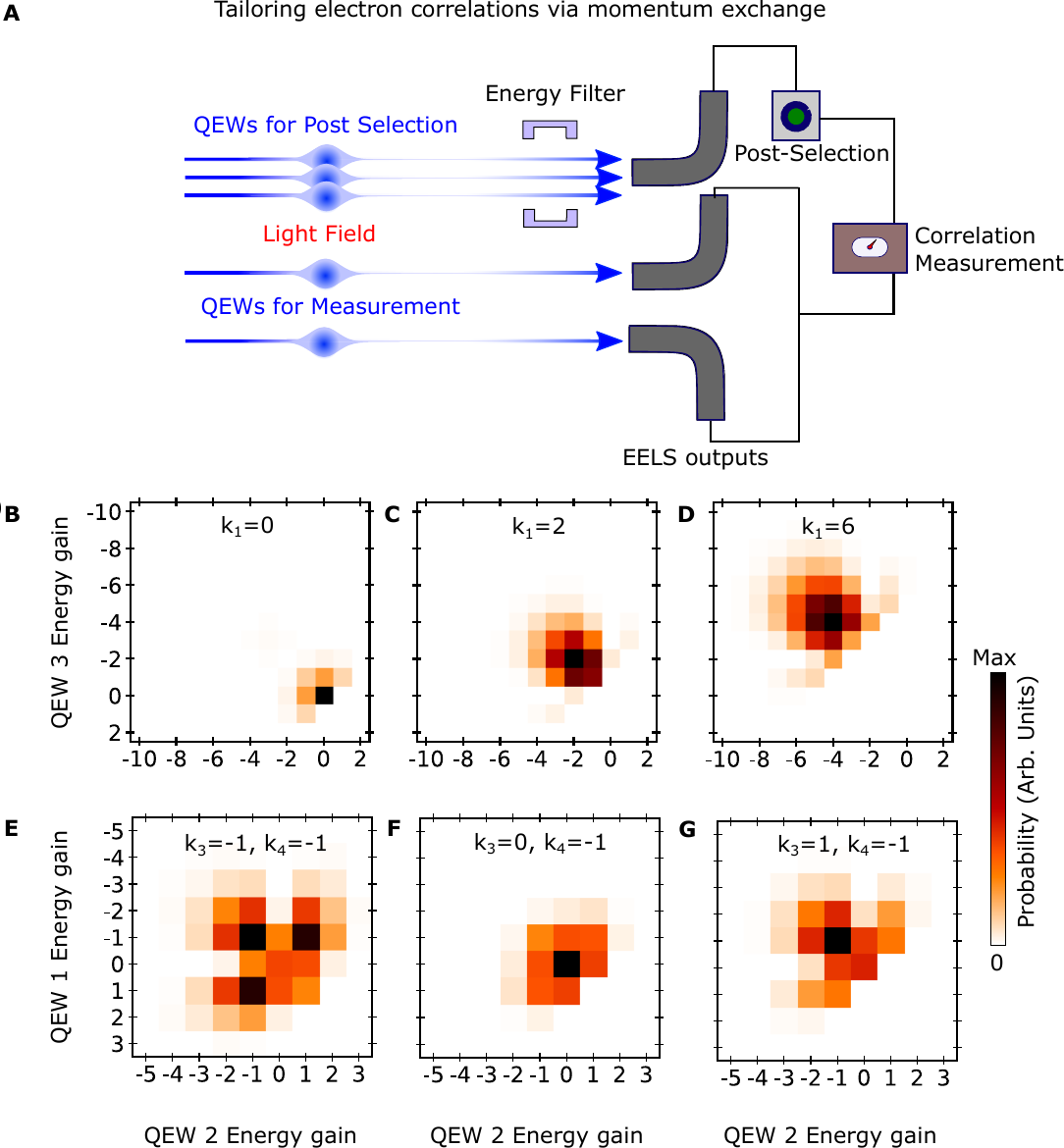}
\caption{\textbf{Tailoring strong correlations between electrons in energy space via post-selection.} (a) We show the schematic of a multi-QEW (Quantum Electron Wavepacket) interaction with inter-electron momentum exchange. We post-select a subset of the QEWs for certain energies. Note that in the case of inter-electron momentum exchange, all electrons are identical as far as the interaction is concerned. We then measure the rest of the QEWs and plot their energy correlations. (b)(c)(d) We plot the correlation plots in the case of inter-electron momentum exchange when we post-select any one of three electrons, i.e. QEW 1 with energy $k_1$ to gain 0, 2 and 6 quanta of energy respectively. We observe that post-selection in the presence of inter-electron momentum exchange can produce highly correlated symmetric correlations between free electrons. Moreover by changing the post-selected energy we can shift the correlations along the symmetry axis of the distribution. The coupling strengths for all cases has been set to $1$ and the initial photonic state is the vaccum Fock state. (e)(f)(g) We post-select any two QEWs (here QEW 3 and QEW 4) for energies $k_3$ and $k_4$ respectively and plot energy correlations between QEW 1 and QEW 2 for $k_{4}=-1, k_{3}=-1,0,1$.}
\label{fig_3}
\end{figure*}
\section*{Results}
We consider an interaction with $N$ QEWs interacting with a single photon mode. We make the assumptions (conventional for PINEM and other free-electron-light interactions) that the electrons are: (i) monoenergetic (unshaped), (ii) travelling through a charge-free region ignoring Coulombic effects and that (iii) their momenta is much stronger than the electromagnetic vector potential $\vec{A}$~\cite{Park2010}. To model an interaction between $N$ QEWs with a single photon mode, we use the interaction Hamiltonian (in the Schrodinger picture, see Methods)
\begin{equation}
    \hat{\mathcal{H}}_{\mathrm{int}}=\sum_{\mu\in\mathbb{Z}^{+}}(g_{\mu}\hat{a}^{\dagger}\hat{b}_{\mu}+g_{\mu}^{*}\hat{a}\hat{b}^{\dagger}_{\mu}),
\label{eq_ham}
\end{equation}
where the subscript $\mu\in Z^{+}$ indicates that the respective quantity is associated with the electron $\mu$ and ${g_{\mu}}$ is the electron-photon coupling strength. The photon ladder operators $\hat{a}$ and $\hat{a}^{\dagger}$ act on Fock states in the usual way: $\hat{a}\ket{n} = \sqrt{n}\ket{n-1}$, $\hat{a}^{\dagger}\ket{n} = \sqrt{n+1}\ket{n+1}$, where $n$ is the number of photons in the Fock state. The ladder operator $\hat{b}_{\mu}$ ($\hat{b}_{\mu}^{\dagger}$) decrements (increments) the electron state $\ket{{j}}$: $\hat{b}\ket{{j}} = \ket{{j-1}}$, $\hat{b}^{\dagger}\ket{{j}} = \ket{{j+1}}$ where an electron in eigenstate $\ket{{j}}$ possesses energy $E_{{j}} = E_{0} + j\hbar\omega_{0}$, $j\in\mathbb{Z}$, $E_{0}$ is the initial electron energy and $\hbar\omega_{0}$ is the photon energy. It should be noted that the electron ladder operators commute with each other: $\left[\hat{b},\hat{b}^{\dagger}\right]=0$.
The corresponding scattering operator (see Methods) is given by:
\begin{equation}
\begin{split}
\hat{S}_{N} =&\ e^{\frac{1}{2}(\sum_{\mu}\mg{{G}_{\mu}}^{2})}e^{-{G}_{N}^{*}\hat{b}_{N}^{\dagger}\hat{a}}\ldots e^{-{G}_{1}^{*}\hat{b}_{1}^{\dagger}\hat{a}}\\
&\times e^{{G}_{1}\hat{b}_{1}\hat{a}^{\dagger}}\ldots
e^{{G}_{N}\hat{b}_{N}\hat{a}^{\dagger}}\\
&\times e^{\frac{1}{2}\sum_{\mu<\nu}({G}_{\mu}{G}_{\nu}^{*}\hat{b}_{\mu}\hat{b}_{\nu}^{\dagger}+h.c.)},
\end{split}
\label{eqn_Smatrix_str_coupling_two_QEWs}
\end{equation}
where $G_{\mu}\equiv -i\frac{g_{\mu}\tau}{\hbar}$ is the dimensionless electron-photon coupling strength term and $\tau$ is the interaction time. In the case of two QEWs, this reduces to:
\begin{equation}
\begin{split}
\hat{{S}}_{2}=&\ e^{\frac{1}{2}(\mg{{G}_{1}}^{2}+\mg{{G}_{2}}^{2})}\overbrace{e^{-{G}_{1}^{*}\hat{b}_{1}^{\dagger}\hat{a}}e^{-{G}_{2}^{*}\hat{b}_{2}^{\dagger}\hat{a}}  
e^{{G}_{1}\hat{b}_{1}\hat{a}^{\dagger}}e^{{G}_{2}\hat{b}_{2}\hat{a}^{\dagger}}}^{\text{PINEM terms}}\\
&\times \underbrace{e^{\frac{1}{2}{G}_{1}{G}_{2}^{*}\hat{b}_{2}^{\dagger}\hat{b}_{1}}
e^{\frac{1}{2}{G}_{1}^{*}{G}_{2}\hat{b}_{2}\hat{b}_{1}^{\dagger}}}_{\text{inter-electron momentum exchange}},
\end{split}
\label{eqn_scop}
\end{equation}
 where $\hat{S}_{2}$ acts on the three-body (one photon and two electrons) state $\ket{n,j,k}$, $n$ is the number of photons, and $j$ ($k$) labels the units of energy gain for electron 1 (electron 2).  For an arbitrary initial state $ \sum_{n_{\text{i}},j_{\text{i}},k_{\text{i}}}C_{n_{\text{i}},j_{\text{i}},k_{\text{i}}}^{(0)}\ket{n_{\text{i}},j_{\text{i}},k_{\text{i}}}$, where $|C_{n_{\text{i}},j_{\text{i}},k_{\text{i}}}^{(0)}|^{2}$ is the probability of finding the initial state in $\ket{n_{\text{i}},j_{\text{i}},k_{\text{i}}}$, and the probability of obtaining a final state $\ket{n_{\text{f}},j_{\text{f}},k_{\text{f}}}$ is given by $P_{n_{\text{f}},j_{\text{f}},k_{\text{f}}} = |\sum_{n_{\text{i}},j_{\text{i}},k_{\text{i}}}C_{n_{\text{i}},j_{\text{i}},k_{\text{i}}}^{(0)}S_{n_{\text{i}},j_{\text{i}},k_{\text{i}}}^{n_{\text{f}},j_{\text{f}},k_{\text{f}}}|^{2}$. Here, $S_{n_{\text{i}},j_{\text{i}},k_{\text{i}}}^{n_{\text{f}},j_{\text{f}},k_{\text{f}}}\equiv \braket{n_{\text{f}},j_{\text{f}},k_{\text{f}}|\hat{S}_{2}|n_{\text{i}},j_{\text{i}},k_{\text{i}}}$ is the S-matrix element (see Eq.~(\ref{eqn_S_matrix_2QEW_final_v1}) in Methods for full expression), and the subscripts ``i'' (``f'') denote the initial (final) quantities. Note that Eq.~(\ref{eqn_scop}) is non-perturbative in the electron-photon coupling strength.

In addition to the usual interaction terms (overbraced terms in Eq.~(\ref{eqn_scop})) found in QPINEM interactions which describe electron-photon energy exchange, it is noteworthy that terms describing electron-electron momentum exchange arise (underbraced terms in Eq.~(\ref{eqn_scop})).  This implies that both electrons can affect each other even in the absence of Coulomb interactions. Physically, the inter-electron momentum exchange terms can be understood as such:  both electrons separately emit and absorb photons, resulting in changes to the photon field itself. In turn, these changes allow both electrons to affect each other.  Since each of the inter-electron momentum exchange terms scale as $|{G}_{1}{G}_{2}|$, we expect the exchange terms to become as significant as the QPINEM terms when $|{G}_{1}| \sim |{G}_{2}| \sim |{G}_{1}{G}_{2}| \sim 1$ -- this is the regime we explore. Physically, this corresponds to the regime in which the electron has a high probability of emitting or absorbing one photon, and so there is also a substantial probability of one electron absorbing/emitting a photon generated by the other electron. We compare the photon and electron distributions for 1-QEW and 2-QEW interactions with quantum light in Supplementary Information (SI) Section I.

In this regime, we can study the effects of inter-electron momentum exchange in interactions between multi-electron pulses and quantum light. In Figure \ref{fig_1} we have considered the case where a two electrons interact with a photon field successively in Figure \ref{fig_1}(a)(b) and then simultaneously in Figure \ref{fig_1}(c)(d). Unlike the successive QEWs, the simultaneous QEWs exchange energy and momentum mediated via the light field. The corresponding joint distribution of final energies of the QEWs is shown for the successive and simultaneous cases in Figure \ref{fig_1}(b) and Figure \ref{fig_1}(d) respectively. We observe a symmetrical joint distribution in electron energies in the case of simultaneous electrons due to inter-electron momentum exchange, resulting in a PCC (see Methods for definition) that is 15 orders of magnitude greater that of the case with successive electrons. Furthermore, the high PCC due to inter-electron momentum exchange is a purely quantum phenomenon in the sense that the high PCC manifests only when light is treated as a quantum object. To see this, we compare the PCC for two-QEW simultaneous interactions when light is a classical field and when light is quantum as shown in Figure \ref{fig_1} (e)(f). We observe that the classical formalism does not generate high PCC. In contrast when the light field is treated as a quantum object, we observe high PCC. For the derivation of the scattering matrix element for a two-QEW interaction with a classical light field, refer to SI Section II. 

Let us now answer the question: Under what conditions would the classical formalism dominate over the quantum formalism, causing inter-electron momentum exchange induced correlations to vanish? We can define the classical limit for a quantum multi-electron interaction as one where the energy exchanged by the electrons is much smaller than the energy of the light field. This corresponds to $N_{p}\gg|G|^2$ where $N_p$ is the number of photons and $G$ is the electron-photon dimensionless coupling. In Figure \ref{fig_2}, we see that for large $N_{p}$ and low $G$, both the case with simultaneous electrons in Figure \ref{fig_2}(a) and the case with successive electrons in Figure \ref{fig_2}(b) correlations approach a similar squarish pattern, which is also what is obtained when performing the simulations with the classical formalism. SI Section III contains a more detailed comparison of the formalisms and how the quantum formalism approaches the classical formalism in the limit of low $N_p$ and large $G$. We also notice that correlation of the QEWs for simultaneous QEW interactions (Figure \ref{fig_2}(a)) is strong for low $N_p$ and large $G$ (Figure \ref{fig_2} (a)).

Apart from getting high PCC as a result of inter-electron momentum exchange between QEWs, we can further tailor correlations by post-selecting one of the participating QEWs. In Figure \ref{fig_3}, we show a schematic to post-select an electron (or more) in a simultaneous interaction between $N$ QEWs and a photon Fock state. As shown in Figure \ref{fig_3} (a), we can post-select QEW 1 to be at a certain energy corresponding to normalized energy gain $k_1$ and then measure the correlation in energy between QEW 2 and QEW 3. Figure \ref{fig_3} (b-d) shows that post-selection of QEW 1 results in highly correlated QEWs whose joint distribution can be modified. For instance, post-selecting for higher QEW 1 energies shifts the distribution in favor of larger photon number and larger QEW energy loss. The resulting pattern of the energy correlation is also distinct from what we can obtain from the 2-QEW cases considered in Figure \ref{fig_2}. In Figure \ref{fig_3} (e-f), we post-select two QEWs out of four QEWs that interact simultaneously with the light field and show the varied patterns that emerge for different post-selected energies. In SI Section IV, we show that the post-selection result remains the same if the energies of two post-selected QEWs are interchanged, which is not a surprise since the QEWs are indistinguishable.
%---------------------------------------------------------
% subsection: Discussion.
%---------------------------------------------------------
\section*{Discussion}  
It is important to note that the high correlations due to field-mediated inter-electron momentum exchange terms have no equivalent in theory which considers a classical field (Supplementary Information (SI) Section I). This is consistent with our theory when we consider an initial photon distribution described a coherent state $\ket{\alpha}$ in the limit $|\alpha|^{2} = n_{\text{av}}\gg 1$ and $|{G}|^{2}\ll 1$ (proof in SI Section II). However, within the strong-coupling regime of interaction of free electrons with quantum light -- currently looked upon as a promising regime in which highly non-classical states of light can be generated, shaped, and probed by free electrons\cite{Baranes2022, BenHayun2021} -- our results show that these inter-electron momentum exchange terms are significant and give rise to substantial enhancements in the correlations (as measured by PCC) between participating QEWs and the QEWs exhibit entanglement i.e., the photon and QEWs are entangled with each other. While we have used the example of Fock state for the light field in all our simulations in the Main Text, SI Section V contains simulations for the cases where the light field is a coherent or thermal light field with the correlations still remaining strong.

Crucially, a recent experiment~\cite{Keramati2021} has shown that two-electron QEWs can already be generated using currently available electron sources~\cite{Hommelhoff2006, Ropers2007,Barwick2007, Yanagisawa2009, Vogelsang2015}. This suggests that our scheme involving two-electron QEWs is already within experimental reach. Advances in electron source technologies, coupled with the results we present here, motivate the study of unprecedented non-Coulombic effects in QPINEM and other free-electron-light interactions involving multi-electron QEWs with quantum states of light.

In the field of quantum computing, the strong non-local entanglement of particles has been studied as part of an effort to realize scalable quantum computation systems that enable parallel quantum computing~\cite{Bluvstein22}. We note that inter-electron momentum exchange is a phenomenon that is non-local compared to say, Coloumb interactions, and can potentially entangle an arbitrarily large number of electrons with a single photon field -- which is not possible with Coulomb interaction due to the repulsive nature of Coulomb forces that limits the number of electrons that can be placed in a limited volume of space. The use of free electrons for quantum computing has several unique advantages over using photon-based or bound-electron based systems. Free electrons readily propagate information across macroscopic distances and are thus a potential candidate for `flying qubits'~\cite{Edlbauer2022, Reinhardt2021}, for instance via the phenomenon of free electron-polariton blockade, made possible by leveraging quantum recoil~\cite{Huang2023, Karnieli_Fan_2023}. These flying qubits are a core component of designing a potential quantum network or quantum internet, which has motivated implementations thus far in semiconductor systems and bound-electron systems~\cite{Edlbauer2022}. The photon-mediated inter-electron momentum exchange mechanism we introduce in this work then empowers the paradigm of using free electrons as flying qubits, by providing a way to correlate multiple flying qubits and potentially to design free-electron based protocols complimenting recent work that have used free-electron-light interactions to implement a free-electron based protocol ~\cite{aviv2023}. In our photon-mediated inter-electron momentum exchange mechanism, the control of the photon mode is crucial as the statistics of the initial photon mode provides an extra degree of freedom in controlling the entanglement. Using free electrons as flying qubits, we could impart information to stationary qubits in the form of photonic states, as has already been shown in the literature~\cite{Dahan2021,BenHayun2021}.

In this work, we have considered examples with up to four electrons, however, our theory accommodates an arbitrary number of electrons interacting with light. We note that there arises a inter-electron momentum exchange term for each electron pair in a multi-QEW interaction. As the number of electrons in a multi-electron pulse increases, Coulombic pair-wise interactions may also increasingly become significant. Recent work\cite{haindl22,Ropers2007} has shown that electron-electron correlations arising from Coulombic interactions can affect shot-noise, electron drift, etc. in the absence of a light field. In the regime where such Coulomb interactions are significant, they will co-exist with inter-electron momentum exchange. In such cases, the relative magnitude of the electron-light coupling to the electron-electron coupling will determine which phenomenon dominates. In low-photon simultaneous interactions with quantum light, we expect electron correlations to favor the electrons losing energy as opposed to the expectation from Coloumbic interactions alone. Above all, we note that the physical origins of Coulomb interactions and inter-electron momentum exchange are very distinct in nature: Coulomb interactions arise from the exchange of virtual photons between QEWs, whereas inter-electron momentum exchange arises from the exchange of physical photons.

Yet another method of tailoring the correlations between participating QEWs could be the use of electron waveshaping. Shaping the QEWs introduces quantum interference\cite{Jeremy2023} which should also co-exist with inter-electron momentum exchange and potentially lead to even more versatility in shaping electron-electron correlations and entanglement compared to the unshaped case. 

In summary, we show that simultaneous interactions involving multi-electron QEWs allow us to tailor electron-electron energy correlations in  ways that go beyond what successive single-electron QEW interactions are capable of. Specifically, the use of multi-electron pulses result in the appearance of inter-electron momentum exchange -- the exchange of energy and momentum between QEWs even in the absence of Coulombic forces. We have shown that inter-electron momentum exchange is a purely quantum phenomenon, manifesting only in regimes where the quantum nature of light is relevant. As a result of inter-electron momentum exchange, outgoing 2-QEW pulses possess PCC over 13 orders of magnitude larger than those obtained when the QEWs interact with the photon mode in succession. We've also shown that inter-electron momentum exchange is a robust means of creating and tailoring strong many-QEW correlations in energy, especially with the use of post-selection techniques. The photon-mediated inter-electron momentum exchange we have studied thus provides an unprecedented means of strongly correlating large numbers of free electrons in energy-momentum space, even if these electrons are spaced far apart enough that Coulombic interactions are no substantial. Our findings fill an important gap in the understanding of multi-electron interaction with photons, and pave the way towards many-electron entanglement techniques for applications like quantum information and ultrafast imaging.
%---------------------------------------------------------
% subsection: METHODS.
%---------------------------------------------------------
\section*{Methods}
\textbf{Physical model for interaction of multiple QEWs with photons} Here, we derive the scattering operator describing the non-perturbative interaction between $N$ paraxial QEWs simultaneously interacting with a common photon mode, as well as the S-matrix element. We assume that all electrons do not interact directly through the Coulomb near-field, which is justified for electron pulses with low electron densities. We also assume there are no other external charges and that the electromagnetic vector potential is much smaller than the electron energies. The full Hamiltonian is given by:
\begin{equation}    \hat{\mathcal{H}}_{\mathrm{total}}=\hat{\mathcal{H}}_{\mathrm{l}}+\hat{\mathcal{H}}_{\mathrm{e}}+\hat{\mathcal{H}}_{\mathrm{int}},
\end{equation}
where $\hat{\mathcal{H}}_{\mathrm{total}}$ is the complete Hamiltonian, $\hat{\mathcal{H}}_{\mathrm{l}}$ is the Hamiltonian term for the light field alone, $\hat{\mathcal{H}}_{\mathrm{e}}$ is the Hamiltonian term for the $N$ electrons and $\hat{\mathcal{H}}_{\mathrm{int}}$ is the interaction Hamiltonian term. The expressions for the light and electrons' Hamiltonian terms in the Schrodinger picture are given by:
\begin{equation}
    \hat{\mathcal{H}}_{\mathrm{l}}=(\hat{a}^{\dagger}\hat{a}+\frac{1}{2})\hbar\omega_{0},\ \hat{\mathcal{H}}_{\mathrm{e}}=\sum_{\mu}\hat{P}_{\mu}v_{\mu},
\end{equation}
where $\hat{a}$($\hat{a}^{\dagger}$) is the annihilation(creation) photon operator satisfying $\left[\hat{a},\hat{a}^{\dagger}\right]=1$, $\omega_{0}$ is the frequency of the light field and, $\hat{P}_{\mu}$ and $v_{\mu}$ are the momentum operator and speed of the electron indexed by $\mu$. The interaction Hamiltonian, which we use to compute the scattering operator, is given by Eq. (\ref{eq_ham}). We obtain the scattering operator from the interaction Hamiltonian as:
\begin{comment}
\begin{equation}
\begin{split}
\hat{{S}_{N}}=e^{(\mathcal{G}_{1}\hat{b}^{\dagger}_{1}\hat{a} - \mathcal{G}^{*}_{1}\hat{b}_{1}\hat{a}^{\dagger})+(\mathcal{G}_{2}\hat{b}^{\dagger}_{2}\hat{a} - \mathcal{G}^{*}_{2}\hat{b}_{2}\hat{a}^{\dagger})+\ldots+(\mathcal{G}_{N}\hat{b}^{\dagger}_{N}\hat{a} - \mathcal{G}^{*}_{N}\hat{b}_{N}\hat{a}^{\dagger})},
\end{split}
\end{equation}
\end{comment}
\begin{equation}
\hat{{S}}=e^{\sum_{\mu=1}^{N}({G}_{\mu}\hat{b}^{\dagger}_{\mu}\hat{a} - {G}^{*}_{\mu}\hat{b}_{\mu}\hat{a}^{\dagger})}.
\end{equation}
Using the Baker-Campbell-Hausdorff formula and separating the hermitian conjugate terms, we obtain Eq. (\ref{eqn_Smatrix_str_coupling_two_QEWs}). Reducing this to the case of two electrons ($N=2$), we recover Eq.~(\ref{eqn_scop}).  Note that inter-electron momentum exchange occurs between both electrons despite no interaction between them due to Coulombic forces. The inter-electron momentum exchange factors arise as a direct result of the non-vanishing commutator $[\hat{a},\hat{a}^{\dagger}]= 1$, and implies that the inter-electron momentum exchange is mediated by the photon field. We now assume that the two electrons QEW 1 and QEW 2 are mono-energetic with initial energies $j_{\text{i}}\hbar\omega$ and $k_{\text{i}}\hbar\omega$  respectively with $j_{\text{i}}$ and $k_{\text{i}}$ being the quantum numbers defining the electron energies. In the examples used in the text, we also assume that the light field is initially in a Fock state with number of photons given by $n_{i}$. The initial state of the three-body system is thus given by $\ket{n_{\text{i}},j_{\text{i}},k_{\text{i}}}$. Similarly we indicate the final state of the system as $\ket{n_{\text{f}},j_{\text{f}},k_{\text{f}}}$. We define the S-matrix element used to calculate the transition probability between the initial and final states to be:

\begin{equation}
S_{n_{\text{i}},j_{\text{i}},k_{\text{i}}}^{n_{\text{f}},j_{\text{f}},k_{\text{f}}} \equiv \braket{n_{\text{f}},j_{\text{f}},k_{\text{f}}|\hat{{S}}_{2}|n_{\text{i}},j_{\text{i}},k_{\text{i}}},
\end{equation}
where $n_{i}$ is the initial number of 
which we solve by expanding the exponential operators in their series expansions. We obtain
\begin{equation}
\begin{split}
S_{n_{\text{i}},j_{\text{i}},k_{\text{i}}}^{n_{\text{f}},j_{\text{f}},k_{\text{f}}} =\ & e^{\frac{1}{2}(|{G}_{1}|^{2}+|{G}_{2}|^{2})} \\
&\times\sum_{l,m,p,q,r,s}\sqrt{\frac{(n_{\text{i}}+l+q)!(n_{\text{f}}+m+p)!}{n_{\text{i}}!n_{\text{f}}!}}\\
&\times\frac{{G}_{1}^{q}{G}_{2}^{l}(-{G}_{1}^{*})^{m}(-{G}_{2}^{*})^{p}}{l!m!p!q!}\frac{({G}_{1}{G}_{2}^{*})^{r}({G}_{1}^{*}{G}_{2})^{s}}{r!s!2^{r+s}} \\
&\times\braket{n_{\text{f}}+m+p|n_{\text{i}}+l+q}\\
&\times\braket{j_{\text{f}}-m+r,k_{\text{f}}-p-r|j_{\text{i}}-q+s,k_{\text{i}}-l-s},
\end{split}
\end{equation}
where integers $l,m,p,q,r,s\in \left[0,\ldots, \infty \right)$. Here, we have split the inner product on the three-body space as the product of two other inner products on the QEWs' space and photon space respectively for legibility. The inner products are Kronecker deltas which enforce the following conditions:
\begin{equation}
\begin{split}
&\Delta n + m + p = l + q,\\
&\Delta k - p - r = -l-s,\\
&\Delta j - m + r = -q + s.
\end{split}
\label{eqn_relations}
\end{equation}

Here, $\Delta k \equiv k_{\text{f}} - k_{\text{i}}$, $\Delta j \equiv j_{\text{f}} - j_{\text{i}}$, and $\Delta n \equiv n_{\text{f}} - n_{\text{i}}$ are the differences between the initial and final quantum numbers. Using all 3 relations gives us the energy conservation relation 
\begin{equation}
\Delta n + \Delta j + \Delta k = 0.
\label{eqn_with_inj_qpinem_COE}
\end{equation}
From the first two conditions in Eq. (~\ref{eqn_relations})  all dependencies on indices $q$ and $l$ vanish, allowing us to obtain the final expression for the S-matrix element:
\begin{equation}
\begin{split}
S_{n_{\text{i}},j_{\text{i}},k_{\text{i}}}^{n_{\text{f}},j_{\text{f}},k_{\text{f}}} =&\ e^{\frac{1}{2}(|{G}_{1}|^{2}+|{G}_{2}|^{2})} \frac{{G}_{1}^{-\Delta j}{G}_{2}^{-\Delta k}}{\sqrt{n_{\text{i}}!n_{\text{f}}!}}\times\\
&\sum_{m,p,r,s}\frac{(-|{G}_{1}|^{2})^{m}(-|{G}_{2}|^{2})^{p}(n_{\text{f}}+p+m )!}{m!p!(s-r+m-\Delta j)!(p - \Delta k -s + r)!}\\
&\frac{|{G}_{1}|^{2s}|{G}_{2}|^{2r}}{r!s!2^{r+s}}.
\end{split}
\label{eqn_S_matrix_2QEW_final_v1}
\end{equation}
Note that since $q,l\geq 0$, it follows that $(p - \Delta k -s + r)\geq 0$ and $(s-r+m-\Delta j)\geq 0 $. For computational efficiency, we find it favorable to use the S-matrix expression with $\hat{a}^{\dagger}$ terms on the LHS and the $\hat{a}$ on the RHS (c.f. Eq.~(\ref{eqn_scop})):
\begin{equation}
\begin{split}
\hat{{S}}_{2} =&\ e^{-\frac{1}{2}|{G}_{1}|^{2}}e^{-\frac{1}{2}|{G}_{2}|^{2}}e^{{G}_{1} \hat{b}_{1}\hat{a}^{\dagger}}
e^{{G}_{2} \hat{b}_{2}\hat{a}^{\dagger}}
e^{-\frac{1}{2}{G}_{1}{G}_{2}^{*}\hat{b}_{2}^{\dagger}\hat{b}_{1}}\\
&\times e^{-\frac{1}{2}{G}_{1}^{*}{G}_{2}\hat{b}_{2}\hat{b}_{1}^{\dagger}}e^{-{G}_{1}^{*} \hat{b}^{\dagger}_{1}\hat{a}} 
e^{-{G}_{2}^{*} \hat{b}^{\dagger}_{2}\hat{a}} .
\end{split}
\label{eqn_Smatrix_str_coupling_two_QEWs_v2}
\end{equation}
Repeating the steps above, we arrive at
\begin{equation}
\begin{split}
S_{n_{\text{i}},j_{\text{i}},k_{\text{i}}}^{n_{\text{f}},j_{\text{f}},k_{\text{f}}} =&\ e^{-\frac{1}{2}(|{G}_{1}|^{2}+|{G}_{2}|^{2})} (-{G}_{1}^{*})^{\Delta j}(-{G}_{2}^{*})^{\Delta k} \sqrt{n_{\text{i}}!n_{\text{f}}!}\\
& \times\sum_{m,p,r,s}\frac{(-|{G}_{1}|^{2})^{s}(-|{G}_{2}|^{2})^{r}}{(n_{\text{f}} - m -p)!r!s!2^{r+s}}\\
&\times\frac{(-|{G}_{1}|^{2})^{m}(-|{G}_{2}|^{2})^{p} }{m!p!}\\
&\times\frac{1}{(\Delta j + m + s - r)!(\Delta k + p + r - s)!},
\end{split}
\label{eqn_S_matrix_2QEW_final_v2}
\end{equation}
where $r,s\in \left[0,\ldots, \infty \right)$, and the other indices have the following bounds:
\begin{equation}
\begin{split}
&0\leq p \leq n_{\text{f}}-m,\\
&0\leq  m\leq n_{\text{f}},\\
&\mathrm{max}\{0,-\Delta j - m+ r\} \leq s \leq \Delta k + p + r.
\end{split}
\end{equation}
The first relation comes from the requirement that $(n_{\text{f}} - m -p) \geq 0$. Since $p\geq 0$, it follows that $n_{\text{f}} - m\geq 0$, which is the second relation. The third relation is obtained by imposing $(\Delta j + m + s - r)\geq 0$ and $(\Delta k + p + r - s)\geq 0$. We see that 3 indices are bounded from above as opposed to Eq.~(\ref{eqn_S_matrix_2QEW_final_v1}), where only $r-s$ is bounded. Hence, Eq.~(\ref{eqn_S_matrix_2QEW_final_v2}) affords us much greater computational efficiency.

We can show that Eq.(~\ref{eqn_S_matrix_2QEW_final_v1}) reduces to the expression for a single-QEW interaction with quantum light\cite{Kfir2019}:
\begin{equation}
\begin{split}
S^{n_{\text{f}},j_{\text{f}}}_{n_{\text{i}},j_{\text{i}}} =& \braket{n_{\text{f}},j_{\text{f}}|\hat{{S}}|n_{\text{i}},j_{\text{i}}} \\
=& \frac{e^{\frac{1}{2}|{G}|^{2}}{G}^{\Delta n}}{\sqrt{n_{\text{f}}!n_{\text{i}}!}}\sum_{m=0}^{\infty}\frac{(-|{G}|^{2})^{m}}{m!(m+\Delta n)!}(n_{\text{f}}+m)!,
\end{split}
\end{equation}
subject to the energy conserving relation
\begin{equation}
\Delta j + \Delta n=0,
\label{eqn_traditiona_qpinem_COE}
\end{equation}
which is consistent with the expression presented in ~\cite{Kfir2019}.\\

As an additional note, the Pearson Correlation Coefficient (PCC) of two QEWs we use in this work is a measure of linear correlation. Given two probability distributions $P(x)$ and $Q(x)$ defined over the same space $X$ where $x\in X$, we define the PCC as:
\begin{equation}
    PCC(P,Q)=\frac{\left\langle{(P-\left\langle{P}\right\rangle)(Q-\left\langle{Q}\right\rangle)}\right\rangle}{\sigma_{P}\sigma_{Q}}
\end{equation}
where $\sigma_{P}$ is the standard deviation of distribution $P$ and $\left\langle P\right\rangle$ is the expectation value of distribution $P$.
\\\\
\begin{acknowledgements}
%%---------------------------------------------------------
%% subsection: DATA AVAILABILITY.
%%---------------------------------------------------------
\textbf{Data availability} The simulation results presented in this paper are available from the corresponding author on reasonable request.\\\\
%
%%---------------------------------------------------------
%% subsection: CODE AVAILABILITY.
%%---------------------------------------------------------
\textbf{Code availability} The code used to compute the results presented in this paper is available from the corresponding author on reasonable request.\\\\
%
%
%%---------------------------------------------------------
%% subsection: author contr..
%%---------------------------------------------------------
\textbf{Author contributions} S.K. and J.L. developed the theory and performed the simulations. L.J.W., J.L. and S.K. conceived the idea and analyzed the results. S.K. and J.L. wrote the manuscript with inputs from L.J.W., N.R., W.W, Y.S.A. and L.K.A. L.J.W. supervised the project.\\\\
%
%%---------------------------------------------------------
%% subsection: compteting interests
%%---------------------------------------------------------
\textbf{Competing interests.} The authors declare no competing interests.\\

%---------------------------------------------------------
% subsection: MISC.
%---------------------------------------------------------

\textbf{Funding}  This project was supported by the Ministry of Education, Singapore, under its AcRF Tier 2 programme (award no. MOE-T2EP50222-0012). LJW acknowledges the Nanyang Assistant Professorship Start-up Grant.  J.L. and L.K.A. acknowledge funding from A*STAR AME IRG (Project ID A2083c0057), MOE PhD Research Scholarship, and USA Office of Naval Research (Global) grant (Project ID N62909-19-1-2047). Y.S.A. is supported by the Singapore Ministry of Education Academic Research Fund Tier 2 (Award No. MOE-T2EP50221-0019). N.R. acknowledges support from a Junior Fellowship from the Harvard Society of Fellows. S.K. acknowledges Sunjun Hwang for his help in proof-reading the manuscript.
\end{acknowledgements}

%---------------------------------------------------------
% subsection: BIB
%---------------------------------------------------------
%\bibliographystyle{apsrev4-1} % uncomment to use bibtex % commentz and use with ``'longbibliography' to include titles

\bibliography{free_electron_cqed}  % uncomment to use bibtex

\end{document}

%\preprint{APS/123-QED}
\title{Supplemental Material\\Strongly correlated multi-electron bunches from interaction with quantum light}% Force line breaks with \\

\author{Suraj Kumar}
\affiliation{
School of Electrical and Electronic Engineering, Nanyang Technological University, 50 Nanyang Avenue, Singapore 639798, Singapore
}

\author{Jeremy Lim}
\affiliation{%
Science, Math and Technology, Singapore University of Technology and Design, 8 Somapah Road, Singapore 487372, Singapore
}%
 
\author{Nicholas Rivera}
\affiliation{Department of Physics, MIT, Cambridge, MA 02139, USA.}

\author{Wesley Wong}
\affiliation{
School of Electrical and Electronic Engineering, Nanyang Technological University, 50 Nanyang Avenue, Singapore 639798, Singapore
}

\author{Yee Sin Ang}
%\email{yeesin_ang@sutd.edu.sg}
\affiliation{%
Science, Math and Technology, Singapore University of Technology and Design, 8 Somapah Road, Singapore 487372, Singapore
}%

\author{Lay Kee Ang}
%\email{ricky_ang@sutd.edu.sg}
\affiliation{%
Science, Math and Technology, Singapore University of Technology and Design, 8 Somapah Road, Singapore 487372, Singapore
}%

\author{Liang Jie Wong}
\email{liangjie.wong@ntu.edu.sg}
\affiliation{
School of Electrical and Electronic Engineering, Nanyang Technological University, 50 Nanyang Avenue, Singapore 639798, Singapore
}

{
\let\clearpage\relax
\maketitle
}

%=======================================================
% APPENDIX: NONPERTURBATIVE DYNAMICS OF DIRAC ELECTRONS
%=======================================================  
\tableofcontents

%=======================================================
% APPENDIX: Classical calculation for two-electron PINEM
%=======================================================  
\section{Comparing photon and electron distributions of free-electron-light interactions involving two-electron QEWs and single-electron QEWs.}
\begin{figure*}[ht!]
\centering
\includegraphics[width = 0.67\textwidth]{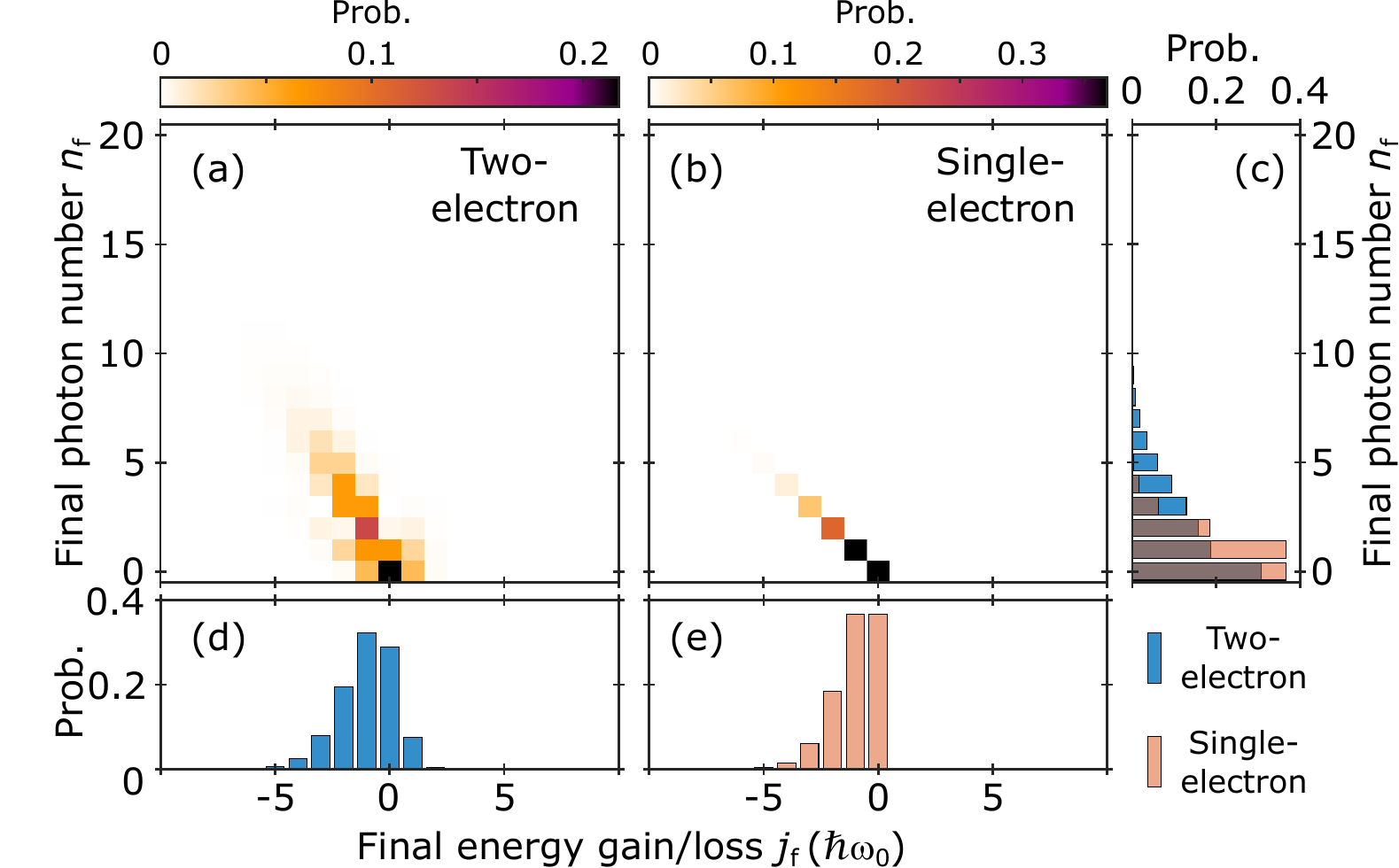}
\caption{Output photon and electron statistics for an initial vacuum Fock state $\ket{n_{\text{i}} = 0}$ and ${G}_{1} = {G}_{2} = 1$, where $n_{\text{i}}$ is the initial photon number and $G_{\mu}$ is the dimensionless photon-electron coupling strength for the electron indexed by $\mu$. Photon-electron joint probability distribution for (a) A free-electron-light interaction with two-electron QEWs, and (b) free-electron-light interaction with single-electron QEW. For (a), we trace over electron 2 states. (c) shows the integrated photon statistics; and (d) and (e) are the integrated output energy gain/loss spectrum for electron 1 corresponding to the two-electron pulse and one-electron pulse scenarios respectively.}
\label{suppfig_for_maintext_fig1_fock_N0_G1p0}
\end{figure*}

\begin{figure}[ht!]
\centering
\includegraphics[width = 0.67\textwidth]{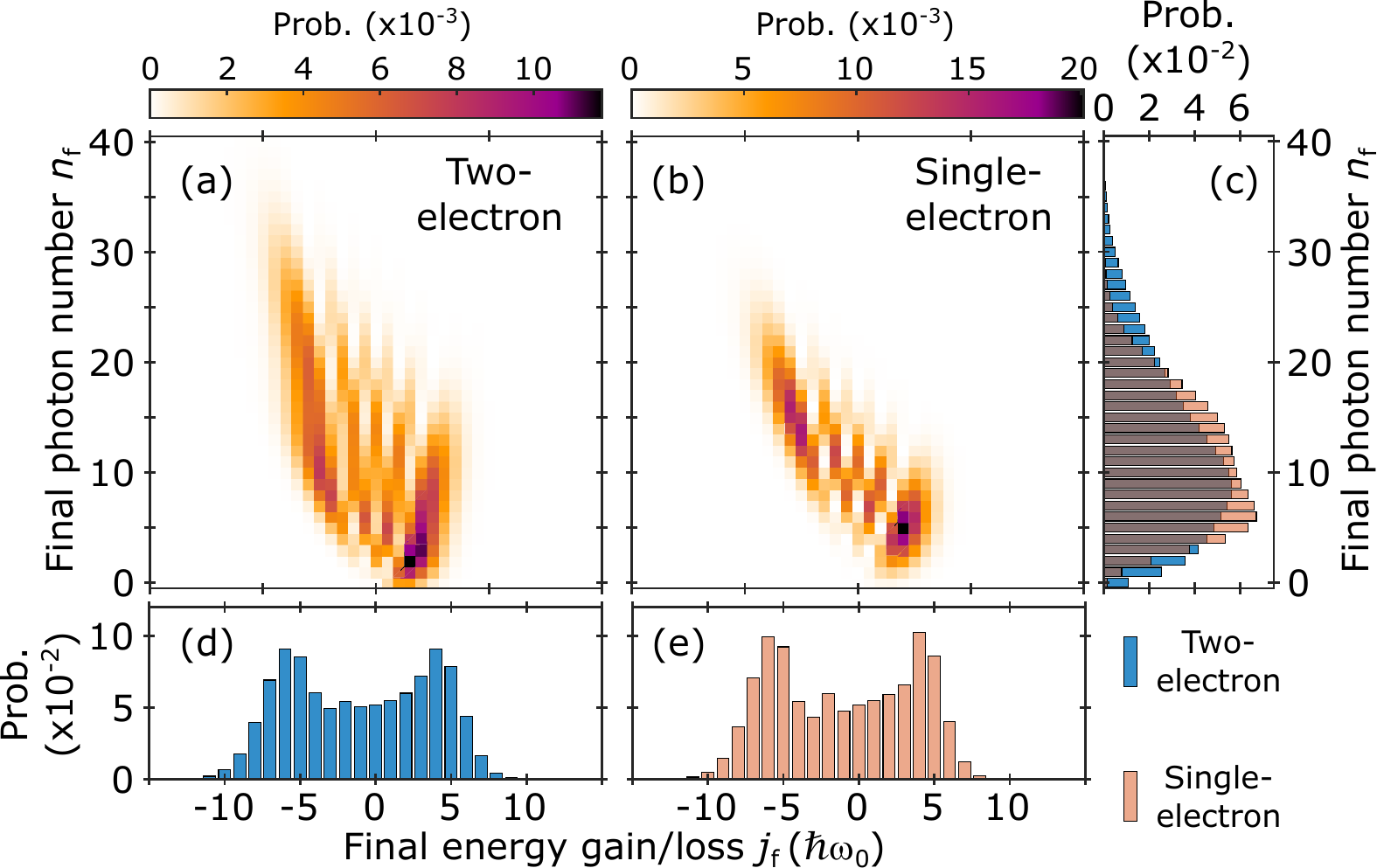}
\caption{Output photon and electron statistics for an initial coherent state with $\langle n\rangle = 10$ and ${G}_{1} = {G}_{2} = 1$. Photon-electron joint probability distribution for (a) Free-electron-light interaction with two-electron QEWs, and (b) Free-electron-light interactions with single-electron QEW. For (a), we trace over electron 2 states. (c) shows the integrated photon statistics; and (d) and (e) are the integrated output energy gain/loss spectrum for electron 1 corresponding to the two-electron QEWs and one-electron QEW scenarios respectively.}
\label{suppfig_for_maintext_fig1_coher_Nav10_G1p0}
\end{figure}

\begin{figure}[ht!]
\centering
\includegraphics[width = 0.67\textwidth]{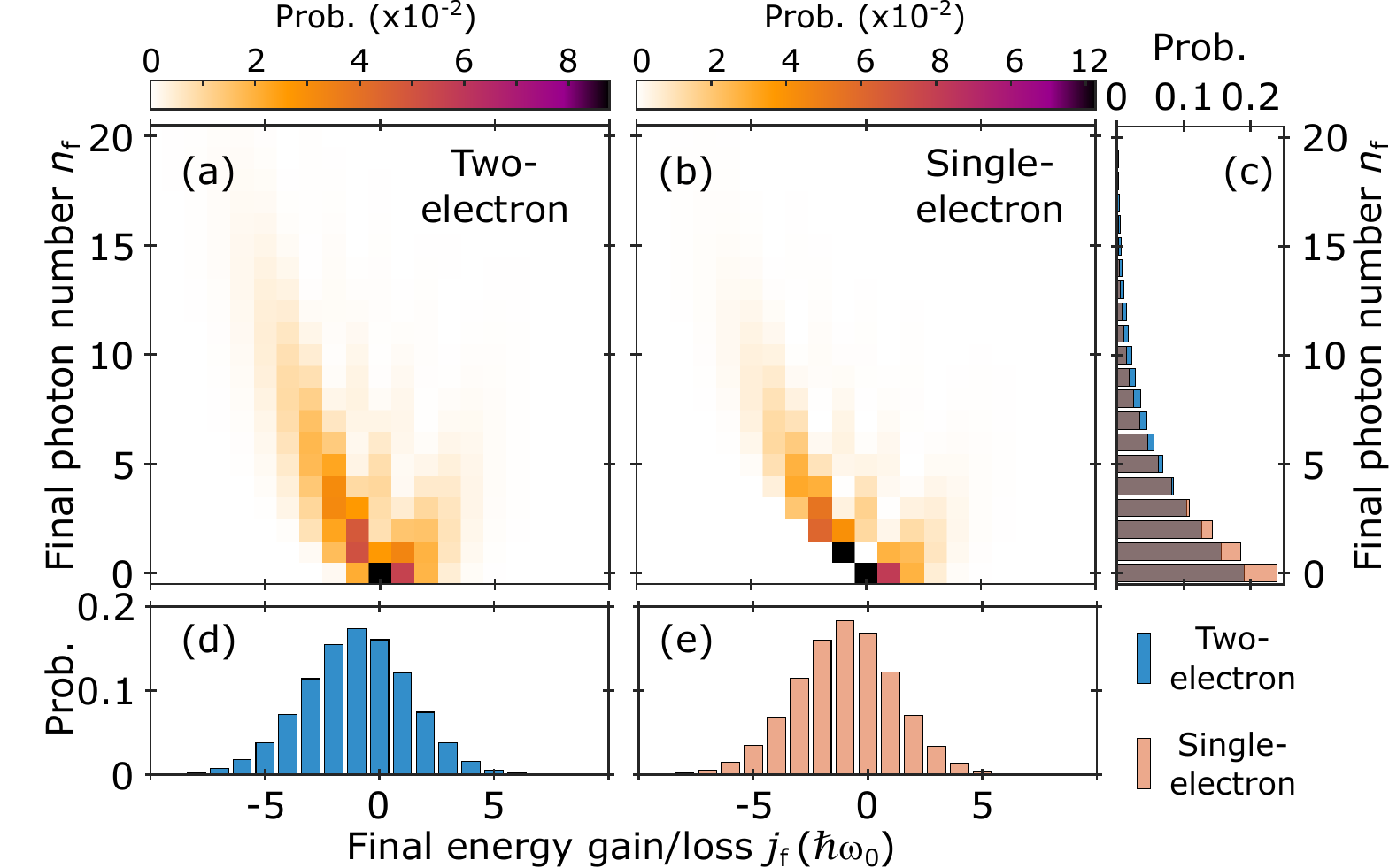}
\caption{Output photon and electron statistics for an initial thermal state with $\langle n\rangle = 2$ and ${G}_{1} = {G}_{2} = 1$. Photon-electron joint probability distribution for (a) Free-electron-light interactions with two-electron QEWs, and (b) Free-electron-light interactions with single-electron QEWs. For (a), we trace over electron 2 states. (c) shows the integrated photon statistics; and (d) and (e) are the integrated output energy gain/loss spectrum for electron 1 corresponding to the two-electron QEWs and one-electron QEWs scenarios respectively.}
\label{suppfig_for_maintext_fig1_thermal_Nav2_G1p0}
\end{figure}

In this section, we consider the final photon and electron statistics and the photon-electron correlations for an initial vacuum Fock state (Fig.~\ref{suppfig_for_maintext_fig1_fock_N0_G1p0}), a coherent state with average photon number $\langle n\rangle = 10$ (Fig.~\ref{suppfig_for_maintext_fig1_coher_Nav10_G1p0}), and a thermal state with average photon number $\langle n\rangle = 2$ (Fig.~\ref{suppfig_for_maintext_fig1_thermal_Nav2_G1p0}).

We find that the photon-electron correlations are generally different for the scenarios involving free-electron-light interactions with two-electron QEWs and single-electron QEWs (panels (a) and (b) respectively in Figs.~\ref{suppfig_for_maintext_fig1_fock_N0_G1p0}-\ref{suppfig_for_maintext_fig1_thermal_Nav2_G1p0}).  We also find that the integrated photon statistics (panel (c)) differ between both cases. The case involving two-electron QEWs generally results in the final photon distribution extending to higher photon numbers as compared to the scenario involving single-electron QEWs. This follows from the different permitted range of attainable final Fock states according to the energy conservation relation for either case: for free-electron-light interactions involving single-electron QEWs, the permitted change in photon number is $\Delta n = -\Delta j$ (Main Text Eq.~(16)), where $\Delta n \equiv n_{\text{f}} - n_{\text{i}}$, and $\Delta j \equiv j_{\text{f}} - j_{\text{i}}$ (Main Text Eq.~(7)) are the changes in photon number and the electron energy (in units of photon energy $\hbar \omega_{0}$). For the interactions with two-electron QEWs, the equivalent expression is $\Delta n = - \Delta j - \Delta k$, where $\Delta k \equiv k_{\text{f}} - k_{\text{i}}$ is the change in the energy of electron 2 in units of $\hbar\omega_{0}$ respectively). Likewise, we find that the integrated electron spectrum differ when two-electron pulses (panel (d)) and single-electron pulses (panel (e)) are involved.

\section{Calculation for two-electron QEW interaction with classical light}
Here, we compute the scattering matrix element for 2 electrons simultaneously interacting with a classical driving field. It is straightforward to generalize the interaction hamiltonian from the single-electron case:
\begin{equation}
\hat{\mathcal{H}}_{\mathrm{int}} = \frac{2\hbar v_{0}}{L}\sin(\omega_{0} t)\bigg{[} \mathcal{G}_{1}\hat{b}_{1}^{\dagger} + \mathcal{G}_{1}^{*}\hat{b}_{1} + \mathcal{G}_{2}\hat{b}_{2}^{\dagger}+ \mathcal{G}_{2}^{*}\hat{b}_{2} \bigg{]}
\end{equation}
where $v_0$ is the electrons' speed, $\omega_{0}$ is the frequency of light, $L$ is the interaction length, $\mathcal{G}_{j}$ is the dimensionless classical coupling strength of the $j$-the electron, where $j\in\{1,2\}$ (not to be confused with the quantum coupling strength ${G}$) and $\hat{b}_{j}$($\hat{b}^{\dagger}_{j}$) is the lowering(raising) operator for the energy of the electron indexed by $j$. $\mathcal{G}_{j}$ is given by the expression:
\begin{equation}
\mathcal{G}_{j}(\Delta k) = \frac{e}{2\hbar\omega}\int^{+L/2}_{-L/2}E_{z}(z)e^{-i\Delta kz}dz,
\end{equation}
where  $z$ is the coordinate along the propagation of the electron, $E_{z}$ is the electric field component along the propagation axis and $\Delta k$ is the change in energy of the electron in units of $\hbar\omega_{0}$.
Casting the operators into the interaction picture, $\hat{b}(t)\rightarrow \hat{b}e^{-i\omega_{0} t}$,  $\hat{b}^{\dagger}(t)\rightarrow \hat{b}^{\dagger}e^{+i\omega_{0} t}$ and dropping all terms that oscillate faster than $\pm\omega_{0}$, the scattering matrix can be written as
\begin{equation}
\hat{{S}} \equiv e^{-\mathcal{G}_{1}^{*}\hat{b}_{1}}e^{-\mathcal{G}_{2}^{*}\hat{b}_{2}}e^{\mathcal{G}_{1}\hat{b}_{1}^{\dagger}}e^{\mathcal{G}_{2}\hat{b}_{2}^{\dagger}},
\end{equation} 
It is straightforward to show that the S-matrix element that connects two incoming monoenergetic electrons, given by the input state $\ket{0,0}$, to the output state $\ket{j_{f},k_{f}}$, where $j_{f}, k_{f}$ are the final energies of the electrons in units of $\hbar\omega_{0}$, has the following form:
\begin{equation}
C_{0,0}^{j_{f},k_{f}} = \braket{j_{f},k_{f}|\hat{{S}}|0,0} = e^{ij_{f}\mathrm{Arg}(\mathcal{G}_{1})}e^{ik_{f}\mathrm{Arg}(\mathcal{G}_{2})} J_{j_{f}}(2|\mathcal{G}_{1}|)J_{k_{f}}(2|\mathcal{G}_{2}|).
\label{eqn_classical_PINEM_2_QEWs}
\end{equation}
Crucially, this tells us that in the limit of non-quantized fields, the joint probability distribution is simply the product of the individual classical probabilities: i.e., the momentum states of both electrons are independent. Thus, the effects of inter-electron momentum exchange are washed out in the limit of non-quantized light.

%=======================================================
% APPENDIX: Two-electron QPINEM in the classical limit
%=======================================================  
\section{{Free-electron-light interactions with quantized and non-quantized light}}
In this section, we show that Eq.~(\ref{eqn_classical_PINEM_2_QEWs}) can be retrieved from our two-electron interaction with quantum light theory in the limit of classical driving fields. A classical light field acts as a reservoir of photons i.e., the number of photons exchanged in the interaction has to be much lesser than the average number of photons. We begin by considering an input state given by 
\begin{equation}
\ket{\alpha,0,0} =\sum_{j=0}^{\infty}e^{-|\alpha|^{2}/2}\frac{\alpha^{j}}{\sqrt{j!}}\ket{j,0,0},
\end{equation}
where the light field is in a coherent state with average photon number $\langle n\rangle = |\alpha|^{2}$.  Using Main Text Eq.~(8), we have
\begin{equation}
\begin{split}
&\braket{{j_{f}},{k_{f}},n_{f}|\hat{{S}_{2}}|0,0,\alpha}\\
=&\sum_{j=0}^{\infty}e^{-|\alpha|^{2}/2}\frac{\alpha^{j}}{\sqrt{j!}}e^{\frac{1}{2}(|{G}_{1}|^{2}+|{G}_{2}|^{2})}\sum_{l,m,p,r,s}\frac{{G}_{1}^{q}{G}_{2}^{l}(-{G}_{1}^{*})^{m}(-{G}_{2}^{*})^{p}}{l!m!p!q!}\frac{({G}_{1}{G}_{2}^{*})^{r}({G}_{1}^{*}{G}_{2})^{s}}{r!s!2^{r}2^{s}}\\
&\times \frac{(j+l+q)!}{\sqrt{j!n_{f}!}}\delta_{j_{f}-m+r,-q+s}\delta_{k_{f}-p-r,-l-s}\delta_{n_{f}+m+p,j+l+q}.
\end{split}
\end{equation}
Enforcing the Kronecker deltas, $m$ and $p$ can be expressed in terms of $q$ and $l$. The inner summand then simplifies to
\begin{equation}
\begin{split}
&\frac{1}{\sqrt{j!}}\sum_{l,m,p,q,r,s}\frac{{G}_{1}^{q}{G}_{2}^{l}(-{G}_{1}^{*})^{m}(-{G}_{2}^{*})^{p}}{l!m!p!q!}\frac{({G}_{1}{G}_{2}^{*})^{r}({G}_{1}^{*}{G}_{2})^{s}}{r!s!2^{r}2^{s}} \frac{(j+l+q)!}{\sqrt{j!n_{f}!}}\\
=&\frac{(-{G}_{1}^{*})^{j_{f}}(-{G}_{2}^{*})^{k_{f}}}{\sqrt{n_{f}!}}\sum_{l,q,r,s}\frac{(-|{G}_{1}|^{2})^{q}(-|{G}_{2}|^{2})^{l}}{q!l!(q+r-s+j_{f})!(l-r+s+k_{f})!}\frac{|{G}_{1}|^{2s}|{G}_{2}|^{2r}}{r!s!2^{r+s}}\frac{(j+l+q)!}{j!}
\end{split}
\end{equation}
Using $j = n_{f}+j_{f}+k_{f}$, we can do away with the sum over $j$:
\begin{equation}
\begin{split}
&\braket{{j_{f}},{k_{f}},n_{f}|\hat{{S}}|0,0,\alpha}\\
=& e^{\frac{1}{2}(|{G}_{1}|^{2}+|{G}_{2}|^{2}-|\alpha|^{2})}(-{G}_{1}^{*}\alpha)^{j_{f}}(-{G}_{2}^{*}\alpha)^{k_{f}}\frac{\alpha^{n_{f}}}{\sqrt{n_{f}!}}\\
&\times\sum_{l,q,r,s}\frac{(-|{G}_{1}|^{2})^{q}(-|{G}_{2}|^{2})^{l}}{q!l!(q+r-s+j_{f})!(l-r+s+k_{f})!}\frac{|{G}_{1}|^{2s}|{G}_{2}|^{2r}}{r!s!2^{r+s}}\frac{(n_{f}+j_{f}+k_{f}+l+q)!}{(n_{f}+j_{f}+k_{f})!}.
\end{split}
\end{equation}
For typical free-electron-light interactions in the classical regime, $|{G}|^{2}\ll 1$, and the electrons only exchange a few quanta of energy with the driving field. Hence, we have $n_{f}\gg k_{f},j_{f}\gtrsim l,q,r,s$ and $\langle j \rangle = |\alpha|^{2} = \langle n_{f} + k_{f} + j_{f}\rangle$. For $(n_{f}+k_{f}+j_{f})\gg l+q$, we can invoke Stirling's approximation:
\begin{equation}
(n_{f}+k_{f}+j_{f}+l+q)!\approx (n_{f}+k_{f}+j_{f})!(n_{f}+k_{f}+j_{f})^{l+q},
\end{equation}
which gives us
\begin{equation}
\begin{split}
&\braket{{j_{f}},{k_{f}},n_{f}|\hat{{S}}|0,0,\alpha}\\
\approx& e^{\frac{1}{2}(|{G}_{1}|^{2}+|{G}_{2}|^{2}-|\alpha|^{2})}(-{G}_{1}^{*}\alpha)^{j_{f}}(-{G}_{2}^{*}\alpha)^{k_{f}}\frac{\alpha^{n_{f}}}{\sqrt{n_{f}!}} \sum_{l,q,r,s}\frac{(-|{G}_{1}\alpha|^{2})^{q}(-|
{G}_{2}\alpha|^{2})^{l}}{q!l!(q+r-s+j_{f})!(l-r+s+k_{f})!}\frac{|{G}_{1}|^{2s}|{G}_{2}|^{2r}}{r!s!2^{r+s}},
\end{split}
\end{equation}
where we have used ${G}\sqrt{n_{f}+k_{f}+j_{f}} \approx {G}\langle\sqrt{n_{f}+k_{f}+j_{f}}\rangle\approx {G}\sqrt{\langle n_{f}+k_{f}+j_{f}\rangle} \approx {G}\alpha$. In going to the last approximation, we can assume that the phases of ${G}$ and $\alpha$ are locked. Since $|{G}|^{2}\ll 1$, we have
\begin{equation}
\begin{split}
&\braket{{j_{f}},{k_{f}},n_{f}|\hat{{S}}|0,0,\alpha}\\
\approx& e^{-|\alpha|^{2}/2}(-{G}_{1}^{*}\alpha)^{j_{f}}(-{G}_{2}^{*}\alpha)^{k_{f}}\frac{\alpha^{n_{f}}}{\sqrt{n_{f}!}}\sum_{l,q}\frac{(-|{G}_{1}\alpha|^{2})^{q}(-|{G}_{2}\alpha|^{2})^{l}}{q!l!(q+j_{f})!(l+k_{f})!}\\
=& e^{-|\alpha|^{2}/2}e^{ij_{f}\mathrm{Arg}(-{G}_{1}^{*}\alpha)}e^{ik_{f}\mathrm{Arg}(-{G}_{2}^{*}\alpha)}\frac{\alpha^{n_{f}}}{\sqrt{n_{f}!}}\sum_{l,q}\frac{(-1)^{q}|{G}_{1}\alpha|^{2q+j_{f}}(-1)^{l}|{G}_{2}\alpha|^{2l+k_{f}}}{q!l!(q+j_{f})!(l+k_{f})!},
\end{split}
\end{equation}
where we have retained only the $r = 0$ and $s = 0$ terms, allowing us to remove the dependence on $r$ and $s$. Using the series definition of Bessel functions and noting that $-{G}^{*}={G}$, the above expression reduces to:
\begin{equation}
\begin{split}
C_{j_{f},k_{f},n_{f}} \approx \underbrace{\frac{\alpha^{n_{f}}e^{-|\alpha|^{2}/2}}{\sqrt{n_{f}!}}}_{C_{n_{f}}} \underbrace{e^{ij_{f}\mathrm{Arg}({G}_{1}\alpha)}e^{ik_{f}\mathrm{Arg}({G}_{2}\alpha)}J_{j_{f}}(2|{G}_{1}\alpha|)J_{k_{f}}(2|{G}_{2}\alpha|)}_{C_{j_{f},k_{f}}}.
\end{split}
\label{eqn_qpinem_classical_fields}
\end{equation}
We recover the expression for the two-electron complex scattering amplitude $C_{j_{f},k_{f}}$ presented in Eq.~(\ref{eqn_classical_PINEM_2_QEWs}), and find that the quantum and classical coupling strengths are related via $|{G}\alpha| = |\mathcal{G}|$, which agrees with known results. Once again, the electron-electron joint probability distribution computed using Eq.~(\ref{eqn_qpinem_classical_fields}) implies that the final momentum states of both electrons are independent.

\section{Post-selecting electron correlations in four-QEW interactions with quantum light}
\begin{figure}
    \centering
    \includegraphics[width=0.7\textwidth]{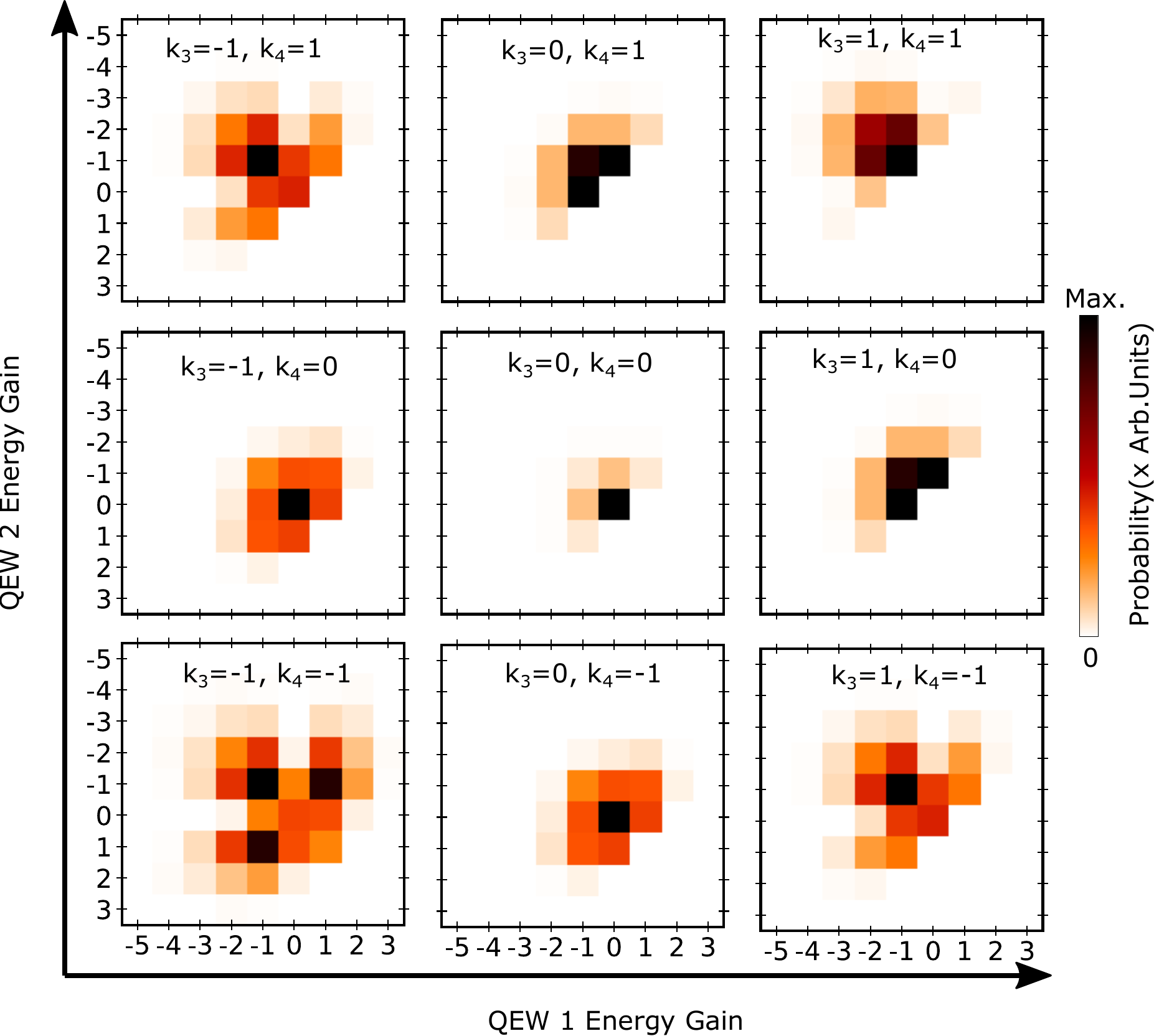}
    \caption{{Post-selecting two-QEW correlations in four-QEW interactions with quantum light.} We consider simultaneous interaction of four QEWs with quantum light and plot the joint probability distribution for the energy gain of QEW 1 vs QEW 2 post-interaction after we post-select QEWs 3 and 4. Here we show the results for QEW 3 and QEW 4 momenta post-selected to be $k_3=-1,0,1$ and $k_4=-1,0,1$ where $k_i$ indicates the final momentum of QEW indexed by $i$.}
    \label{fig2}
\end{figure}

In the Main Text Figure 3, we have considered the example of a three-QEW simultaneous interaction with quantum light wherein we post-select a single QEW to manipulate the correlations between the other two QEWs. We extend the principle to four-QEW interactions with quantum light wherein we post-select two QEWs. In Figure \ref{fig2}, we plot the QEW 1 versus QEW 2 energy gain joint probability distribution map when we post-select QEW 3 and QEW 4 momenta to be $k_3$ and $k_4$ respectively for varying final electron momenta. We observe that the distribution doesn't change if we exchange the values of $k_3$ and $k_4$. For instance $k_3=1,k_4=0$ and $k_3=0,k_4=1$ have the same distribution. Moreover no matter how we post-select the final momenta, the distribution is always symmetrical. These results are the consequences of every QEW being identical with respect to the interaction so exchanging QEW 3 and QEW 4 post-selected momenta preserves the distribution and every distribution observed is symmetrical.

\section{Two-QEW interactions with thermal and coherent states of light}
\begin{figure}
    \centering%
    \includegraphics[width=0.9\textwidth]{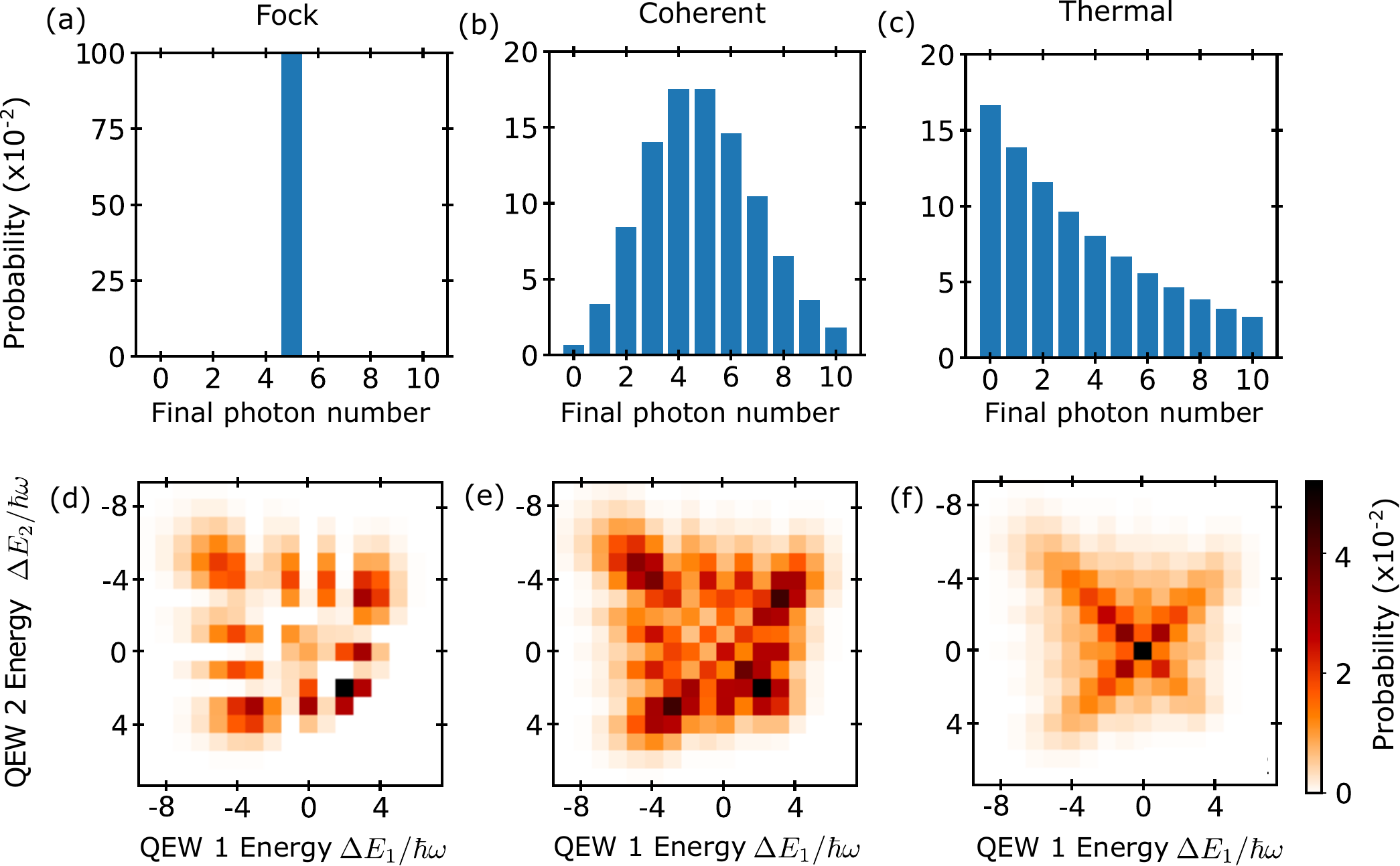}%
    \caption{{Strong electron correlations obtained using two simultaneous electrons interacting with quantum states of light.} We consider simultaneous interaction of two QEWs with quantum light for different initial states of the light field. In (a),(b) and (c) we show the initial probabilities of photon numbers in the initial photon state for Fock, coherent and thermal states of light with average photon number $\langle 
 n\rangle=5$ in all cases. Below in (d)(e)(f) we plot the joint probability distribution for the electron energies for the Fock, coherent and thermal states respectively and the corresponding Pearson coefficient of correlation. In all cases the electron-light coupling constants have been fixed to be $G_1=G_2=1$.}
    \label{fig3}
\end{figure}

Throughout the Main Text, we have considered the light field to be in a Fock state. However our results of higher correlations hold true for other quantum states of the light field such as the coherent state and the thermal state. We define the coherent state and the thermal state photon probabilities $P_{c}$ and $P_{t}$ in relation to the average photon number, $\langle n\rangle$ as:
\begin{equation}
    P_{c}(n)=e^{-\langle n\rangle}\frac{\langle n\rangle^n}{n!}
\end{equation}
\begin{equation}
    P_{t}(n)=\frac{1}{\langle n\rangle+1}\left(\frac{\langle n\rangle}{\langle n\rangle+1}\right)^n
\end{equation}
In Figure \ref{fig3} (a)(b)(c) we show the probability distributions for the Fock, coherent and thermal states of light with average photon number $n_{av}=5$ respectively. Figure \ref{fig3} (d)(e)(f) shows the joint probability distributions for the electron energies post-interaction for the Fock, coherent and thermal states respectively. We observe that strong (relative to the successive case) correlations are observed for all initial states of light. We know from Figure 2 from the Main Text that for simultaneous electrons interacting with quantum light in a Fock state, the quantum mutual information decreases as initial photon number increases. For a more complex initial state such as thermal or coherent state, we can represent the initial state as a superposition of Fock states. Note that the higher the probability of low-number Fock states in the initial state, the higher the correlations post-interaction.
